\documentclass[sigconf, 10pt]{acmart}

\newgeometry{left=1.8cm,right=1.8cm,bottom=1.8cm,top=1.8cm}
\usepackage{booktabs} 
\usepackage{url} 
\usepackage{graphicx} 
\usepackage{amsmath,amssymb,amsfonts} 
\usepackage{xcolor} 
\usepackage[labelfont=bf]{caption} 
\usepackage{textcomp} 
\usepackage{dblfloatfix}\usepackage{url} 
\usepackage{graphicx} 
\usepackage{amsmath,amssymb,amsfonts} 
\usepackage{xcolor} 
\usepackage[labelfont=bf]{caption} 
\usepackage{textcomp} 
\usepackage{dblfloatfix}
\usepackage{anyfontsize}
\usepackage{fancyhdr}

\usepackage{soul} 
\usepackage{enumitem} 
\usepackage{lipsum} 
\usepackage[hang,flushmargin]{footmisc}
\usepackage{etoolbox}
\usepackage{float}
\usepackage{multirow}
\usepackage{hhline}
\newlist{inlinelist}{enumerate*}{1}
\setlist[inlinelist]{label=(\roman*)}

\newcommand{\changed}[1]{\textcolor{black}{#1}}

\usepackage{algorithmicx}
\usepackage{algorithm}
\usepackage{algpseudocode}
\usepackage{wrapfig}

\algnewcommand\algorithmicinput{\textbf{Input:}}
\algnewcommand\Input{\item[\algorithmicinput]}

\algnewcommand\algorithmicconst{\textbf{Constraints:}}
\algnewcommand\Const{\item[\algorithmicconst]}

\algnewcommand\algorithmicoutput{\textbf{Output:}}
\algnewcommand\Output{\item[\algorithmicoutput]}

\algnewcommand{\algorithmicgoto}{\textbf{go to}}%
\algnewcommand{\Goto}[1]{\algorithmicgoto~\ref{#1}}%

\algrenewcommand\algorithmicindent{0.5em}

\usepackage{array}
\newcolumntype{L}[1]{>{\raggedright\let\newline\\\arraybackslash\hspace{0pt}}m{#1}}
\newcolumntype{C}[1]{>{\centering\let\newline\\\arraybackslash\hspace{0pt}}m{#1}}
\newcolumntype{R}[1]{>{\raggedleft\let\newline\\\arraybackslash\hspace{0pt}}m{#1}}
\usepackage{makecell}
\usepackage{diagbox}

\setcopyright{rightsretained}

\acmConference[DAC'19]{ACM Design Automation Conference}{June 2019}{Las Vegas, Nevada, USA}
\acmYear{2019}
\copyrightyear{2019}

\acmArticle{4}
\acmPrice{15.00}

\begin{document}
\settopmatter{printacmref=false} 
\renewcommand\footnotetextcopyrightpermission[1]{} 
\pagestyle{plain} 
\title{XBioSiP: A Methodology for Approximate Bio-Signal Processing at the Edge}

\settopmatter{authorsperrow=1}
\author{Bharath Srinivas Prabakaran, Semeen Rehman, Muhammad Shafique}
\affiliation{%
    \institution{\textit{Vienna University of Technology (TU Wien), Vienna, Austria}}
}
\email{{bharath.prabakaran, semeen.rehman, muhammad.shafique}@tuwien.ac.at}
\pagestyle{fancy}
\lhead{Accepted for publication at the Design Automation Conference 2019 (DAC'19), Las Vegas, Nevada, USA}
\rhead{}
\cfoot{\thepage}


\author{}


\begin{abstract}
Bio-signals exhibit high redundancy, and the algorithms for their processing are inherently error resilient. This property can be leveraged to improve the energy-efficiency of IoT-Edge (wearables) through the emerging trend of approximate computing. This paper presents \textit{XBioSiP}, a novel methodology for approximate bio-signal processing that employs two quality evaluation stages, during the pre-processing and bio-signal processing stages, to determine the approximation parameters. It thereby achieves high energy savings while satisfying the user-determined quality constraint. Our methodology achieves, up to $19\times$ and $22\times$ reduction in the energy consumption of a QRS peak detection algorithm for $0\%$ and $<1\%$ loss in peak detection accuracy, respectively.
\end{abstract}

%
%



\maketitle
\section{Introduction}
\label{sec:Introduction}
IoT-health sensor nodes~\cite{istepanian2011potential}, wearable patches like sweat sensor-arrays~\cite{gao2016fully}, and electronic devices like Fitbit~\cite{diaz2015fitbit}, Apple Watch~\cite{khushhal2017validity}, portable monitors, etc. are used to collect data related to various physiological signals such as skin temperature, heart-rate, pulse and respiration rate, electrocardiogram (ECG), electromyogram (EMG), and electroencephalogram (EEG).
Due to their small form-factor, these edge nodes are severely resource and energy constrained, and might not be suitable to perform highly compute-intensive operations.
Furthermore, these kind of battery-operated sensor nodes are expected to have long lifespans and perform real-time data processing, which consumes a lot of energy.  
Therefore, reduced energy consumption and extreme energy-efficiency are desirable objectives in IoT edge devices.

The re-emerging approximate computing paradigm can be seen as a possible solution for building highly energy-efficient systems.
Error-resilient applications have the ability to produce acceptable output despite having incorrect or approximate intermediate computations~\cite{mittal2016survey}. 
It is due to four major factors: (i) noise and redundancy in the real-world datasets being processed, (ii) error attenuating computational patterns of applications, (iii) varying levels of perception by different users, and (iv) non-existence of a unique golden output for several use-cases~\cite{chippa2013analysis}. 
Prior works have shown such resilience for different applications like pattern recognition, machine learning, big data mining, and image and video processing.
This resilience towards errors can be leveraged to introduce approximations across the computing stack~\cite{shafique2016cross}\cite{venkataramani2015approximate} at various hardware~\cite{gupta2011impact}\cite{gupta2013low}\cite{kulkarni2011trading}\cite{rehman2016architectural} and software layers~\cite{baek2010green}\cite{esmaeilzadeh2012architecture}\cite{laurenzano2016input}\cite{mishra2014iact} to achieve high efficiency gains in terms of area, power, latency, and energy consumption. 

Therefore, this paper aims at answering the following fundamental questions:
\begin{enumerate}[label=(\arabic*),leftmargin=*]
	\item If and how can \textit{approximate computing} be employed to significantly reduce the energy consumption in IoT edge devices for bio-signal processing in healthcare applications?
	\item What should be the impact of such approximations on the output of bio-signal processing applications, which are typically considered sensitive in nature?
\end{enumerate}


Towards this, we make the following \textbf{Novel Contributions:}
\begin{itemize}
	\item We study the error-resilience of bio-signal processing applications using a motivational analysis on ECG processing (see Section~\ref{subsec:motivation}).
	\item We propose \textit{XBioSiP}, a novel two-stage quality eval-uation-based approximation methodology that maximizes the energy-efficiency of bio-signal processing applications while satisfying user-defined quality constraints (see Section~\ref{sec:Methodology}). 
	\item We propose a novel three-phase design generation methodology that traverses the design space and generates possible approximate processing units that offer high energy reductions while satisfying the user-defined quality constraints (see~Section~\ref{subsec:ASFLP}).
	\item We evaluate the efficacy of our novel methodology using an ECG processing application as a case study. We use the Pan-Tompkins algorithm for QRS Peak detection as the target application\cite{pan1985real} (see Section~\ref{subsec:ERA}).
\end{itemize}

We illustrate the area, latency, power and energy reductions obtained by synthesizing the approximate processing unit using an ASIC tool-flow. \textit{We achieve energy reductions of \textasciitilde$19.7\times$ with $0\%$ loss in peak detection accuracy} (see Section~\ref{sec:Results}).
The RTL and behavioral models of these approximate adders and multipliers, including a VDHL implementation of the key stages present in the Pan-Tompkins algorithm are released as an open-source library at \textcolor{blue}{\url{https://xbiosip.sourceforge.io/}}.
This will help facilitate researchers to reproduce our work and enable further research and development in this field.
\section{Motivational Analysis}
\label{subsec:motivation}

\noindent \changed{In this section, first, we analyze the total energy consumption of the bio-signal monitoring sensor nodes to study the potential for minimizing overall energy consumption.
These sensor nodes are responsible for performing three main functions, namely, 
\begin{inlinelist}
    \item sensing and collecting the bio-signal data,
    \item processing (targeted region) the real-world data, and
    \item communicating the data to the next network layer for long-term storage.
\end{inlinelist}
Fig.~\ref{fig:InitAnalysis} illustrates the sensing energy and total energy consumption of five bio-signal monitoring nodes (adapted from studies presented in~\cite{nia2015energy}). 
As can be observed, the sensing energy is at least six orders of magnitude less than the total energy consumed by the device.
Furthermore, previous studies have shown that on-sensor processing energy constitutes of $40\%-60\%$ of the total energy consumed by the sensor nodes~\cite{rault2014energy}.
In this work, we focus on minimizing the energy consumption of on-sensor processing in order to extend the device lifetime.
Next, we analyze the error-resilience of a target application and the energy reductions obtained by leveraging the application's inherent error-resilience.}


\setcounter{figure}{0}
\begin{figure}[h]
	\centering
	\captionsetup{justification=raggedright,singlelinecheck=false}
	\includegraphics[width = \linewidth]{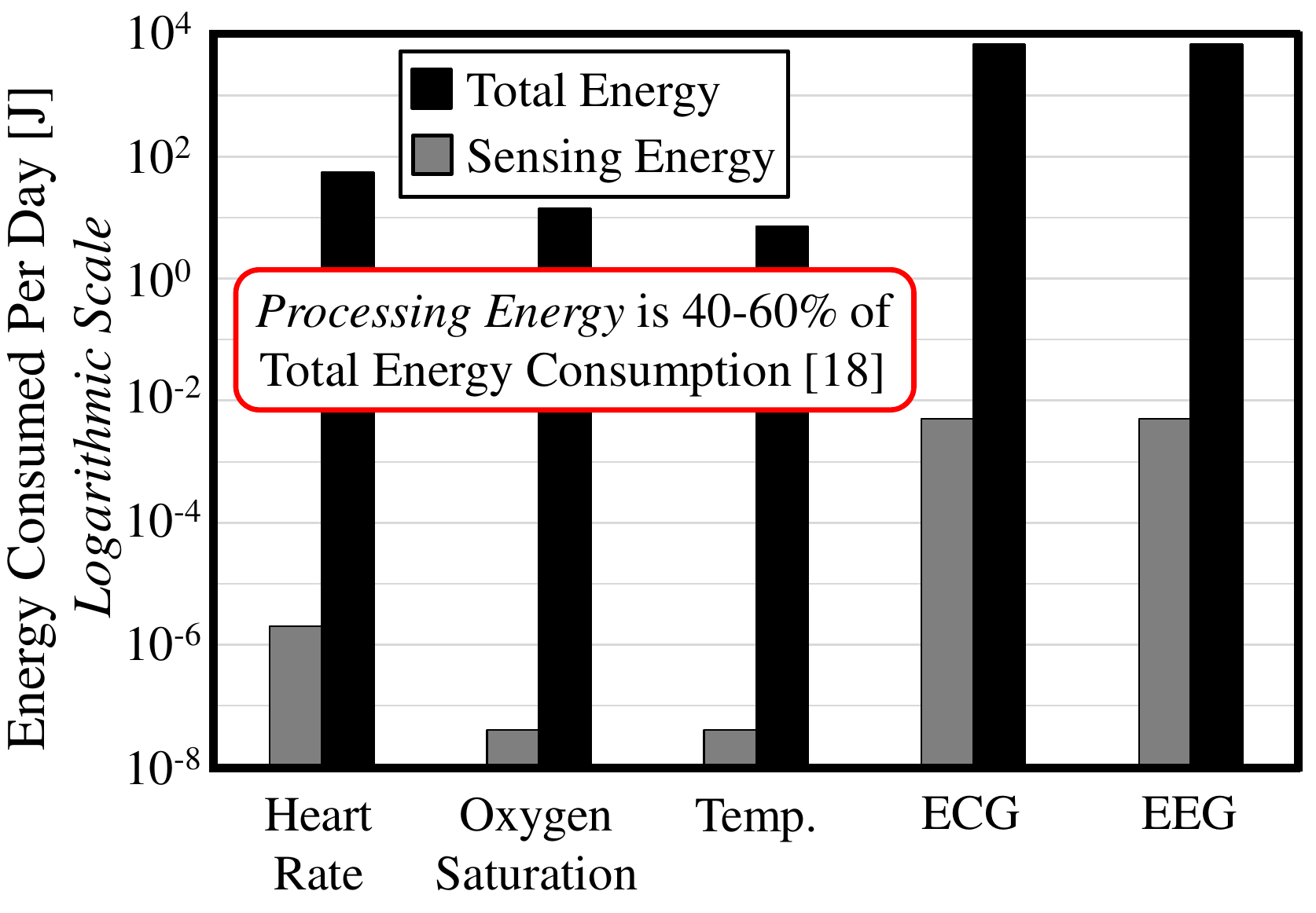}
	\caption{\textbf{Energy Consumption of Five Bio-signal Measuring Sensor Nodes (adapted from~\cite{nia2015energy}\cite{rault2014energy}).}}
	\label{fig:InitAnalysis}
\end{figure}

\textbf{Target Application:} Physiological bio-signals like heart-rate, blood pressure, skin temperature, ECG, EEG, EMG, etc. are analog in nature and are converted to the digital domain for storage and processing. 
These bio-signals are noisy and permeated with sparsity and redundancy. 
Furthermore, the digital signal processing applications used to filter and extract viable information from this data are inherently error resilient due to their computational patterns that deploy filters and aggregators, which attenuate errors. 
Hence, these signal-processing applications are amenable to approximations that can be exploited to increase the energy efficiency of the system by approximating the underlying arithmetic blocks like adders and multipliers used in these applications.
To understand this error-resilience behavior of bio-signal processing systems and the potential of approximate computing, we perform the following experimental study on the Pan-Tompkins algorithm~\cite{pan1985real}.

\textbf{Case Study: Low Pass Filter.} First, we analyze the initial stage of the Pan-Tompkins QRS peak detection algorithm (presented in Section~\ref{sec:Background}) to quantify and evaluate its error resilience. 
It employs a 10th order, 11-tap, Low Pass Filter (LPF) that comprises 10 adders, 11 multipliers, and 10 registers. 
In this work, we focus on approximating the arithmetic operators (i.e., the adders and multipliers) of the filter, to reduce energy consumption (energy reductions). 
In this experiment, we use the low-power approximate 1-bit full-adder, Approximate Adder-5, proposed by Gupta \textit{et al.}~\cite{gupta2011impact}~\cite{gupta2013low} and the approximate elementary $2\times2$-multiplier module proposed by Rehman \textit{et al.}~\cite{rehman2016architectural}, along with their accurate counterparts, to construct higher bit-width approximate adder and multiplier blocks.
An overview of the elementary adder and multiplier modules, along with their area and power properties, is presented in Section~\ref{subsec:AppAddMult}. 

The results of this experiment are illustrated in Fig.~\ref{fig:Motivation}.
On the \textit{x}-axis, we denote the number of output LSBs approximated in the LPF. 
Note, the number of LSBs approximated decides which of the computationally accurate 1-bit full-adder and elementary $2\times2$ multiplier modules are replaced with their approximate counterparts. 
The \textit{y}-axes denote the achieved reductions in terms of area, latency, power, and energy compared to the accurate case, as well as the peak detection accuracy and output signal quality (in terms
of structural similarity index: SSIM).

\setcounter{figure}{1}
\begin{figure}[h]
	\centering
	\captionsetup{justification=raggedright,singlelinecheck=false}
	\includegraphics[width = \linewidth]{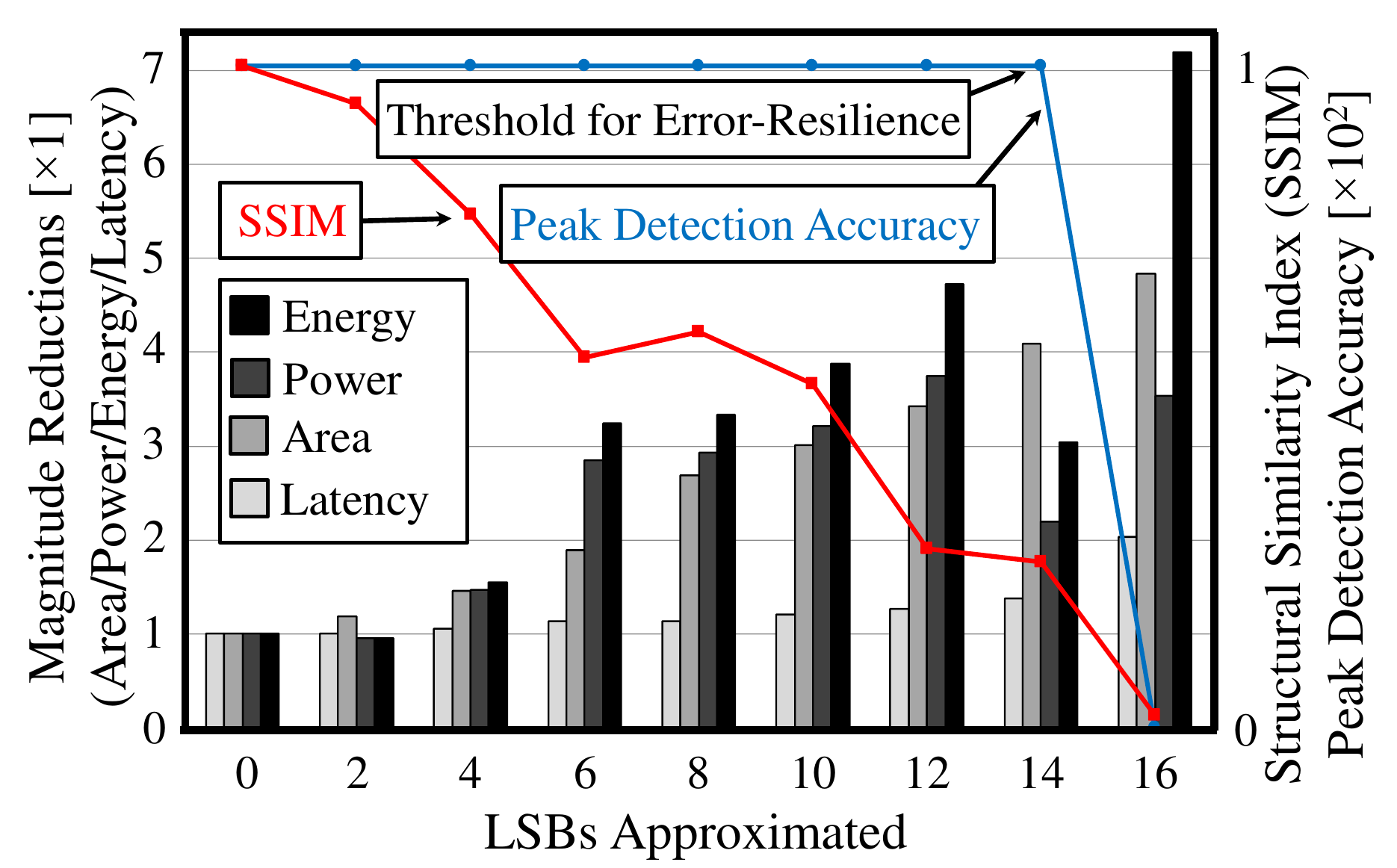}
	\caption{\textbf{Error Resilience of the Low Pass Filter Stage.}}
	\label{fig:Motivation}
\end{figure}

\setcounter{figure}{2}
\begin{figure*}[!b]
	\centering
	\captionsetup{justification=raggedright,singlelinecheck=false}
	\includegraphics[width = \linewidth]{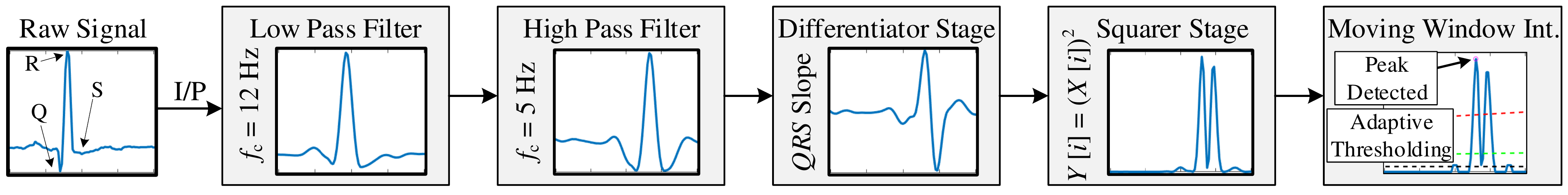}
	\caption{\textbf{Overview of the Pan-Tompkins QRS Peak Detection Algorithm for ECG Bio-signals.}}
	\label{fig:PanTompkins}
\end{figure*}

From these experiments, we make the following \textbf{\textit{key observations:}}
\begin{itemize}[leftmargin=*]
	\item Increasing the number of approximated LSBs decreases the area, energy, power, and/or latency.
	\item \textit{The peak detection accuracy is consistently $100\%$ for an increasing number of LSBs approximated, exhibiting high tolerance towards intermediate approximation errors.}
	\item The error resilience threshold for this stage is 14 LSBs, after which the peak detection accuracy falls to zero, which implies that the application is significantly resilient to approximation errors.
	\item The output signal quality, obtained after the second stage of the application (used by physicians for health monitoring and diagnosis), drastically decreases when approximating more than 2 LSBs, as illustrated by the SSIM metric. However, if $50\%$ loss in signal quality can be tolerated, we can approximate up to 10 LSBs while achieving \textasciitilde$4\times$ energy reductions.
\end{itemize}

Similarly, the subsequent four stages of the algorithm are error-resilient as well (see Section~\ref{subsec:ERA}), which can be leveraged to further reduce energy consumption at each stage of the bio-signal processing application.
\changed{While previous works have focused on deploying aggressive voltage scaling in memories~\cite{bortolotti2014approximate} to achieve \textasciitilde$5\times$ energy reductions, in this work, we focus on deploying functional approximations in the processing elements, thereby limiting the maximum error.}
Before proceeding to the main technical contributions of this paper, we provide a brief overview of our target application (the Pan-Tompkins algorithm for QRS Peak detection in ECG signals), to the level of detail necessary for understanding our technique.
\section{Background: Pan-Tompkins Algorithm}
\label{sec:Background}
\noindent The QRS peak detection algorithm proposed by Jiapu Pan and Willis Tompkins in 1985~\cite{pan1985real}, still serves as a basic standard for QRS detection in a wide-range of hospital or portable Holter monitors and wearable electronic devices. 
The algorithm is used to determine the number of heartbeats in the sampled ECG signal by detecting the QRS complex present in each correct heartbeat (cardiac arrhythmias produce incorrect ECG waveforms).
The analog ECG signal is sampled at a frequency of 200 Hz, using a 16-bit ADC. The Pan-Tompkins algorithm is composed of five key stages (see Fig.~\ref{fig:PanTompkins}):
\begin{enumerate}[label=\textbf{(\Alph*)}, leftmargin=*]
	\item \textbf{Low Pass Filter:} First, to eliminate high frequency noise due to muscle movement and electrical interference, a low pass filter is used to eliminate frequencies above 12 Hz.
	\item \textbf{High Pass Filter:} Next, to remove low frequency noise components from the input, such as respiration and baseline wander, and obtain signals within the desired range of 5-12 Hz, a high pass filter with a cut-off frequency ($f_c$) of 5 Hz is used.
	\item \textbf{Differentiator Stage:} After the initial data pre-processing and noise filtering, a five-tap digital differentiator is used to determine the QRS complex slope information.
	\item \textbf{Squarer Stage:} The signal is then squared point-by-point, which nonlinearly amplifies the output while emphasizing the higher (ECG) frequencies and renders all data points positive.
	\item \textbf{Moving Window Integration:} Finally, to extract the waveform feature information and the slope of the R-wave (see Fig.~\ref{fig:PanTompkins}), a moving average filter is deployed.
\end{enumerate}
In the next section, we describe \textit{XBioSiP}, our novel methodology for approximating bio-signal processing at the edge.
\section{Our XBioSip Methodology for Approximate Bio-Signal Processing}
\label{sec:Methodology}

\noindent An overview of our methodology for approximating arithmetic blocks in bio-signal processing algorithms is presented in Fig.~\ref{fig:Methodology}.
It is composed of four key steps, starting with the design and evaluation of the elementary (approximate) arithmetic modules.
Next, we analyze the error-resilience of each stage in the application by varying the number of LSBs approximated to extract its energy-quality trade-off.
Using the previous analyses we propose a design generation methodology that explores a limited number of points in the design space to generate approximate designs that offer maximum energy reductions while satisfying the user-defined quality constraints.

\setcounter{figure}{3}
\begin{figure}[t]
	\centering
	\captionsetup{justification=raggedright,singlelinecheck=false}
	\includegraphics[width = \linewidth]{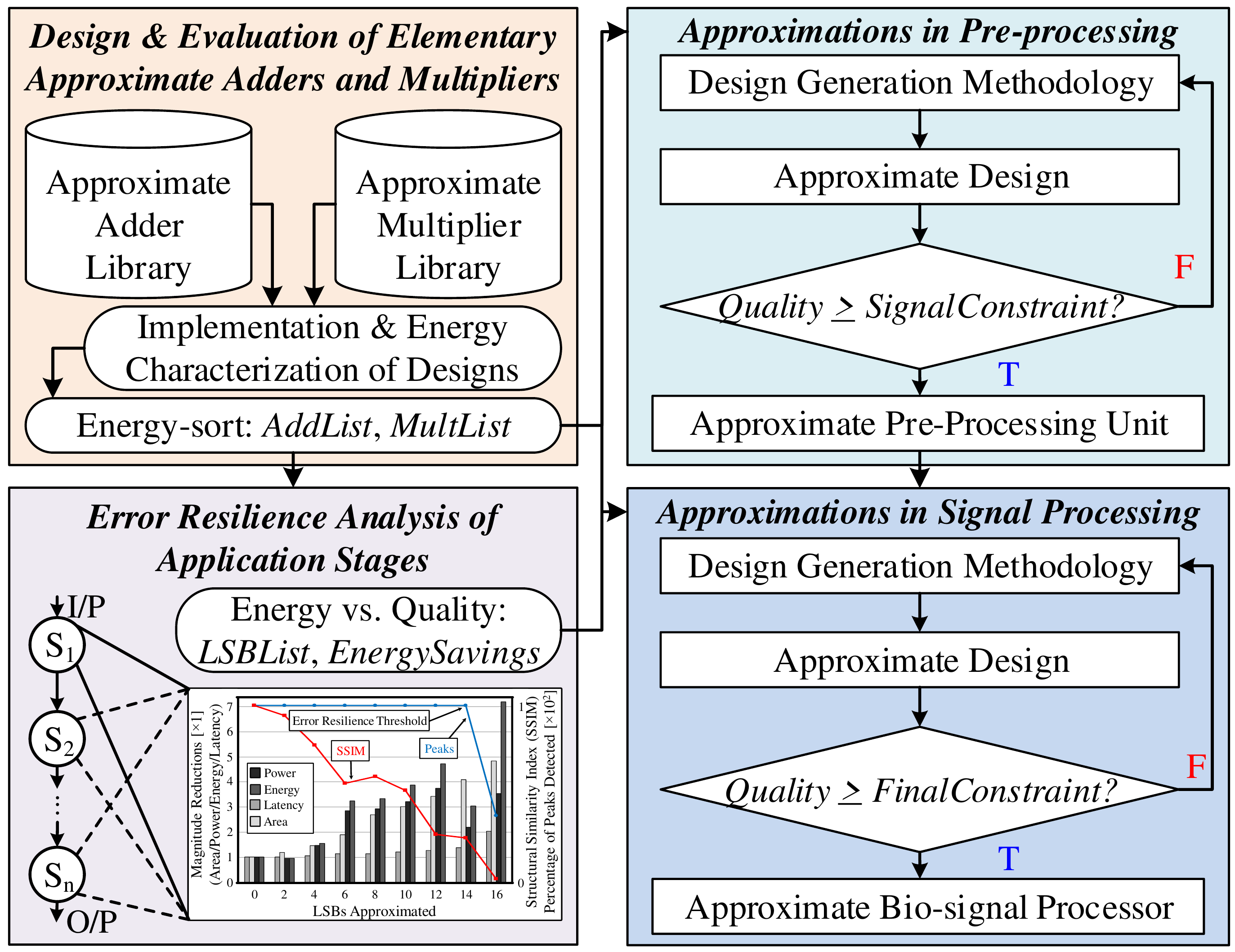}
	\caption{\textbf{Overview of \textit{XBioSiP} Methodology.}}
	\label{fig:Methodology}
\end{figure}

Bio-signal processing algorithms are essentially composed of two sections.
The first section is data pre-processing, which comprises techniques for signal reconstruction, noise filtering, data transformation, reduction and compression, etc. 
The second section is responsible for efficiently extracting the relevant information from the pre-processed data.
We utilize the proposed design generation methodology in the data pre-processing and the main bio-signal processing sections to generate approximate processing units. 
We propose to evaluate the quality of output signals at two stages to ensure fine-grained quality-control, and to provide a medical physician access to the user's accurate health history in case of emergencies.
The key difference between the two stages is the use of a different quality metric for evaluating the constraint. 
The output obtained from data pre-processing is generally in the form of a signal whose quality can be estimated using metrics like PSNR and/or SSIM, whereas the final metric depends on the output of the bio-signal application targeted for approximation.
In our case, the Pan-Tompkins algorithm is used to estimate the number of QRS peaks in the signal because of which we consider peak detection accuracy as the final metric.


\setcounter{figure}{7}
\begin{figure*}[b]
    \vspace{0.05cm}
	\centering
	\captionsetup{justification=raggedright,singlelinecheck=false}
	\includegraphics[width = \linewidth]{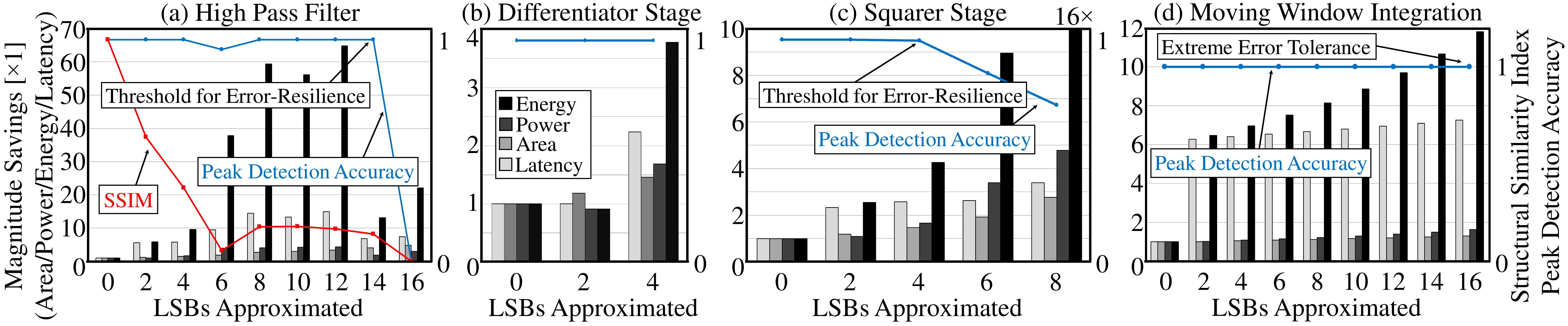}
	\caption{\textbf{Error Resilience Analysis of Pan-Tompkins Application Stages.}}
	\label{fig:ERA}
\end{figure*}

\subsection{Elementary Approximate Adders and Multipliers}
\label{subsec:AppAddMult}

\noindent In this paper, we use the low-power approximate 1-bit full-adders (FAs) proposed by Gupta \textit{et al.}~\cite{gupta2011impact}~\cite{gupta2013low} and the approximate $2\times2$ multiplier modules proposed by Kulkarni \textit{et al.}~\cite{kulkarni2011trading} and Rehman \textit{et al.}~\cite{rehman2016architectural}.
This library of elementary approximate arithmetic modules, along with their accurate counterparts, is illustrated in Fig.~\ref{fig:AddMult}.
\setcounter{figure}{4}
\begin{figure}[t]
	\centering
	\captionsetup{justification=raggedright,singlelinecheck=false}
	\includegraphics[width = \linewidth]{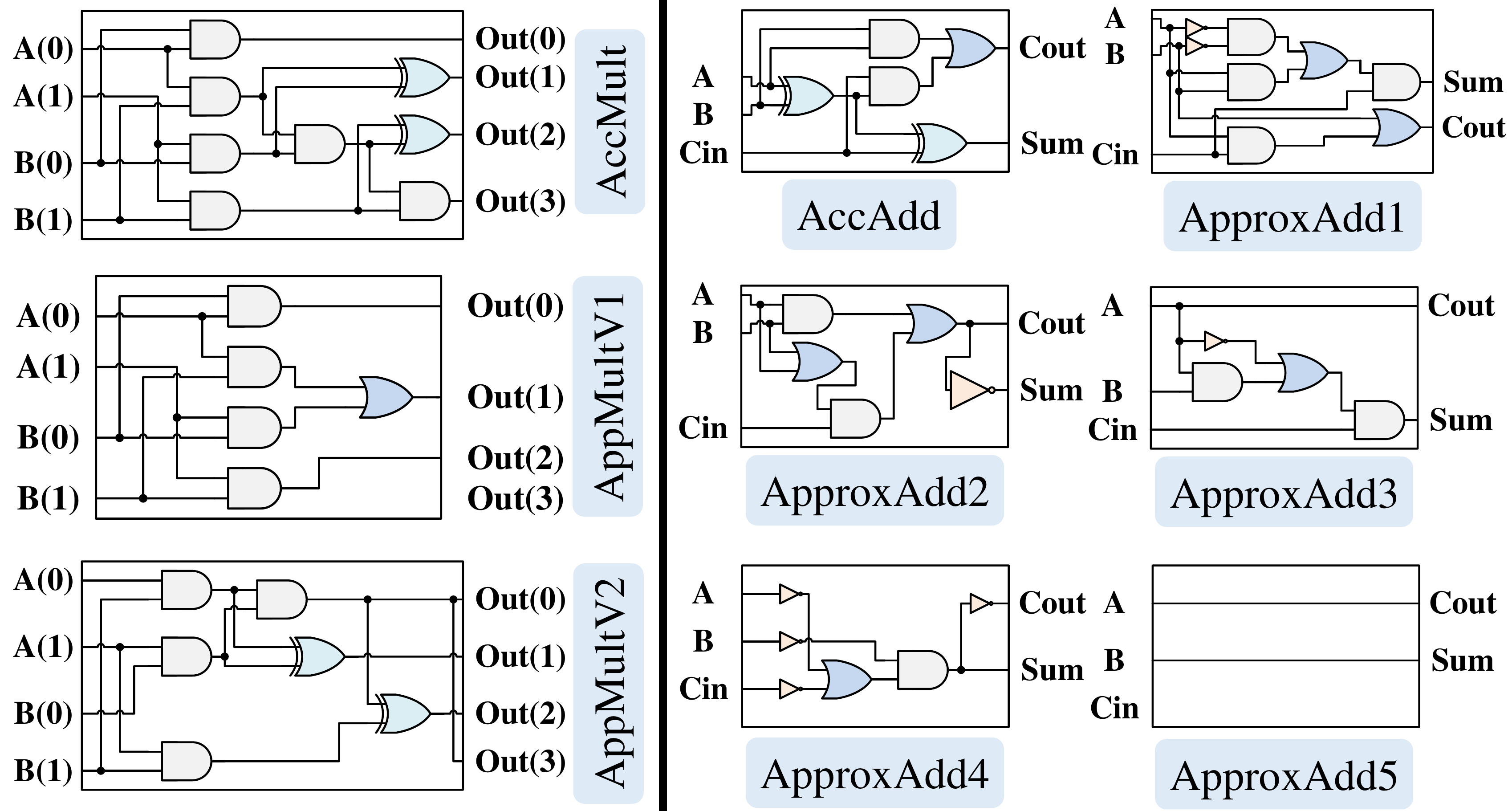}
	\caption{\textbf{Approximate Multipliers and Adders}~\textbf{\cite{gupta2011impact}\cite{gupta2013low}\cite{kulkarni2011trading}\cite{rehman2016architectural}.}}
	\label{fig:AddMult}
\end{figure}
These arithmetic modules are used to construct the larger bit-width approximate adder and multiplier blocks.
Fig.~\ref{fig:Adder} presents a ripple-carry adder architecture that is used to construct larger bit-width approximate adders by replacing the accurate 1-bit FA modules with approximate ones.
We restrict ourselves to deploying approximations at the LSBs to limit the error magnitude.
Similarly, larger bit-width multiplier blocks are recursively constructed from elementary multipliers and adders, similar to the structure shown in Fig.~\ref{fig:Multiplier}. 
For example, a $16\times16$ multiplier is recursively partitioned into four smaller $8\times8$ multiplier blocks, whose outputs are accumulated using three $32-$bit adders.
Similarly, each $8\times8$ multiplier is sub-partitioned into four smaller $4\times4$ multiplier blocks, each of which is further sub-partitioned into four elementary $2\times2$ multipliers.

\begin{figure}[h]
	\centering
	\captionsetup{justification=raggedright,singlelinecheck=false}
	\includegraphics[width = 0.7\linewidth]{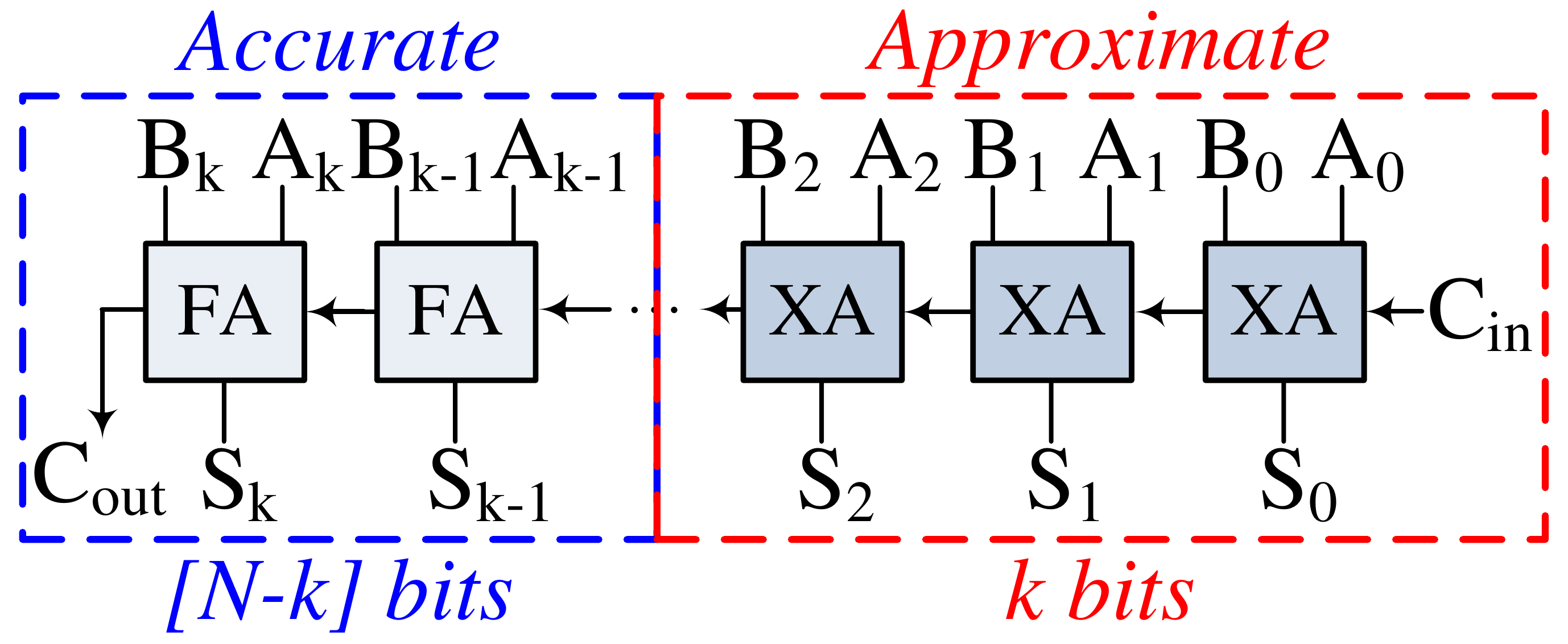}
	\caption{\textbf{Approximate Larger Bit-Width Ripple Carry Adder.}}
	\label{fig:Adder}
\end{figure}
\begin{figure}[h]
	\centering
	\captionsetup{justification=raggedright,singlelinecheck=false}
	\includegraphics[width = \linewidth]{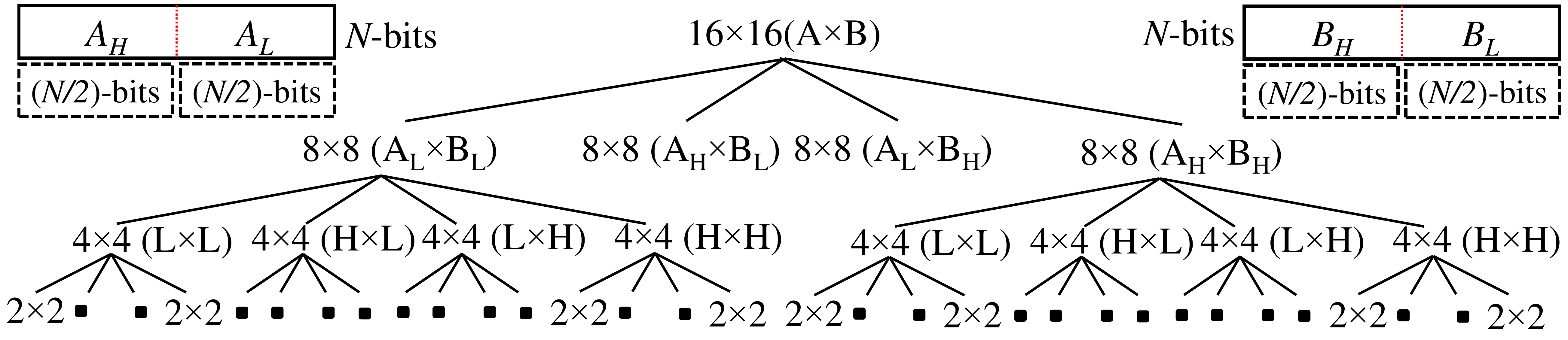}
	\caption{\textbf{Larger Bit-Width Recursive Multiplier Designs.}}
	\label{fig:Multiplier}
\end{figure}

\vspace{0.1cm}
\begin{table}[t]
	\centering
	\caption{\protect\centering \textbf{Synthesis Results of Our Elementary Approximate Adder and Multiplier Library.}}
	\begin{tabular}{L{1cm}|C{1cm}|C{1cm}|C{1cm}|C{1.2cm}|}
		\cline{2-5}
		& Area [$\mu m^2$] & Delay [$ns$] & Power [$\mu W$] & Energy [$fJ$] \\ \hline
		\multicolumn{1}{|l|}{Accurate}   & 10.08     & 0.18         & 2.27      & 0.409       \\ \hline
		\multicolumn{1}{|l|}{ApproxAdd1} & 8.28      & 0.11         & 1.34      & 0.147       \\ \hline
		\multicolumn{1}{|l|}{ApproxAdd2} & 3.96      & 0.08         & 0.61      & 0.049       \\ \hline
		\multicolumn{1}{|l|}{ApproxAdd3} & 3.60      & 0.06         & 0.41      & 0.025       \\ \hline
		\multicolumn{1}{|l|}{ApproxAdd4} & 3.24      & 0.06         & 0.33      & 0.020       \\ \hline
		\multicolumn{1}{|l|}{ApproxAdd5} & 0.00      & 0.00         & 0.00      & 0.000       \\ \hline
	\end{tabular}
	\label{table1}
\end{table}
\vspace{0.1cm}
\begin{table}[t]
	\centering
	\begin{tabular}{L{1.77cm}|C{1cm}|C{1cm}|C{1cm}|C{1.2cm}|}
		\cline{2-5}
		& Area [$\mu m^2$] & Delay [$ns$] & Power [$\mu W$] & Energy [$fJ$] \\ \hline
		\multicolumn{1}{|l|}{Accurate}   & 14.40    & 0.16       & 1.80      & 0.288       \\ \hline
		\multicolumn{1}{|l|}{AppMultV1} & 11.52    & 0.13       & 1.67      & 0.167       \\ \hline
		\multicolumn{1}{|l|}{AppMultV2} & 9.72     & 0.06       & 1.37      & 0.137       \\ \hline
	\end{tabular}
\end{table}


To obtain the area, latency, power, and energy properties, we synthesize these arithmetic modules using an ASIC tool-flow (see Section~\ref{sec:ES}). 
The corresponding results are presented in table~\ref{table1}.
The elementary arithmetic modules are listed in descending order of energy consumption, which is considered while generating the approximate designs as our primary goal is to maximize the energy reductions.


\subsection{Error Resilience Analysis of Application Stages}
\label{subsec:ERA}

\noindent In this step, we quantify the error resilience of each application stage by varying the number of LSBs approximated with the least energy consuming arithmetic adder and multiplier modules.
We analyze the energy-quality trade-offs obtained for each application stage, similar to the low-pass filter analysis presented in Section~\ref{subsec:motivation}.
Fig.~\ref{fig:ERA} (a)-(d) illustrate the error resilience and energy reductions of the remaining stages in the application for a varying number of LSBs approximated. 
We make the following key observations for each application stage:
\begin{itemize}[leftmargin=*]
	\item \textit{Low Pass Filter:} By applying approximations at this stage, \textasciitilde$5\times$ energy reductions with $0\%$ loss in peak detection accuracy can be achieved, when up to $14$ LSBs are approximated, as shown in Fig.~\ref{fig:Motivation}. Approximating up to 8 LSBs can also achieve energy reductions of \textasciitilde$3\times$ while tolerating less than $50\%$ loss in output signal quality (SSIM).
	\item \textit{High Pass Filter:} Due to the number of adders and multipliers in this stage (31 and 32, respectively), approximating just 8 LSBs can achieve energy reductions of \textasciitilde$60\times$, with $0\%$ loss in peak detection accuracy, as shown in Fig.~\ref{fig:ERA}(a). However, the output signal quality (SSIM) drastically decreases after approximating merely 2 LSBs.
	\item \textit{Differentiator:} The magnitude of filter-coefficients at this stage is very small (2 and 1), and approximating more than 4 LSBs truncates all active paths, effectively connecting the outputs to either the inputs or to logic `0'. Moreover, due to the importance of this stage in extracting the QRS slope information, and the limited number of adders and multipliers, applying approximations in this stage is ineffective and leads to limited energy reductions, as shown in Fig.~\ref{fig:ERA}(b).
	\item \textit{Squarer:}
	This stage only requires a large bit-width multiplier. Hence, approximating just a few bits can lead to significant degradation in output quality as illustrated in Fig.~\ref{fig:ERA}(c). Therefore, the approximation potential for this stage is low.
	\item \textit{Moving Window Integration:}
	The final stage is composed solely of adder blocks and, as illustrated by Fig.~\ref{fig:ERA}(d), is extremely error-resilient, tolerating approximations of up to 16 LSBs, while achieving \textasciitilde$12\times$ energy reductions.
\end{itemize}

\subsection{Approximations in Data Pre-Processing and Signal Processing}
\label{subsec:ASFLP}

\noindent Using the initial analyses and energy reports available from the earlier steps, we propose a novel design generation methodology that effectively explores the design space to generate designs that offer high energy reductions while satisfying the user-defined quality constraints.
The pseudo-code of our design generation methodology has been presented in Algorithm~\ref{Algo1}.
It is used to generate potential approximate processing units for the target application.
All the information gathered from the previous stages, i.e., the number of elementary adders and multipliers, maximum number of LSBs that can be approximated at each stage, quality loss, energy reductions obtained for a varying number of approximate LSBs, etc. are provided as an input to the design generation methodology.

The first phase of the methodology starts with initializing two empty arrays to store the designs that satisfy the quality constraint. 
Afterwards, the application stages currently being approximated are sorted in an ascending order, based on the maximum energy reductions obtained at each individual stage (line~\ref{alpha}).
Next, we start evaluating possible approximations in the first stage present in \textit{StageList}, starting from the maximum possible number of LSBs that can be approximated in the given stage, using the least energy-consuming adder and multiplier modules.
We construct a behavioral model of the current stage given the approximation parameters, \textit{LSB, Mult,} and \textit{Add}, to evaluate the output quality of the current design.
If the quality constraint is satisfied by the current approximate design, we store the current design as a potential approximate architecture for the current application stage (lines~\ref{Marker1}$-$\ref{Marker1b}).

Note, our main goal is to find an approximate design that offers maximum energy reductions while satisfying the quality constraint. 
This is achieved by evaluating limited number of points in the design space to reduce the exploration time.
\begin{algorithm}[H]
    \normalsize
	\caption{Design Generation Methodology}
	\label{Algo1}
	\begin{algorithmic}[1]
		\Input $\{EnergySavings, LSBList\}~\forall~Stages, StageList$
		\Input $AddList, MultList$
		\Const $QualConst.$
		\Output $Stages[Architecture]$
		\State $Stage1 = Array[];$
		\State $Stage2 = Array[];$
		\State $AscendingSort(StageList, EnergySavings);$\label{alpha}
		\State $Stage = StageList(1);$\Comment{Initialize First Phase}\label{Marker1}
		\For{\texttt{LSB \textbf{in} Stage[LSBList]}}
		\For{\texttt{Mult \textbf{in} MultList}}
		\For{\texttt{Add \textbf{in} AddList}}
		\State $Design = Stage[Architecture,LSB, Mult, Add];$\par
		\State $OutputQuality = Evaluate(Design);$\par
		\If{$OutputQuality \geq QualConst.$}
		\State $Stage1.append(Design);$
		\State \Goto{Marker2}
		\EndIf
		\EndFor
		\EndFor
		\EndFor\label{Marker1b}
		\Comment{Second Phase}
		\For{$i$ $\gets$ $2$ $to$ $size(StageList)$} \label{Marker2}
		\State $Stage = StageList(i);$
		\For{\texttt{LSB \textbf{in} Reverse(Stage[LSBList])}}
		\For{\texttt{Mult \textbf{in} Reverse(MultList)}}
		\For{\texttt{Add \textbf{in} Reverse(AddList)}}
		\State $Design = Stage[Architecture,LSB, Mult, Add];$
		\State $OutputQuality = Evaluate(Design);$
		\If{$OutputQuality < QualConst.$}
		\State \Goto{Marker3}
		\Else
		\State $Stage2.append(Design);$
		\EndIf
		\EndFor
		\EndFor
		\EndFor\label{Marker2b}
		\Comment{Third Phase}
		\While{\texttt{StageList(i-1)[Architecture,LSB]$\geq$2}} \label{Marker3}
		\For {\texttt{Mult \textbf{in} MultList}}
		\For {\texttt{Add \textbf{in} AddList}}
		\State $LSB1 = StageList(i-1)[Architecture, LSB] - 2;$
		\State $LSB2 = StageList(i)[Architecture, LSB] + 2;$
		\State $Design1 =$\par 
		\hskip\algorithmicindent $StageList(i-1)[Architecture, LSB1, Mult, Add];$
		\State $Design2 = $\par
		\hskip\algorithmicindent $StageList(i)[Architecture,LSB2, Mult, Add];$
		\State $OutputQuality = Evaluate(Design1, Design2);$
		\If{$OutputQuality \geq QualConst.$}
		\State $Stage1.append(Design1);$
		\State $Stage2.append(Design2);$
		\EndIf
		\EndFor
		\EndFor
		\EndWhile\label{Marker3b}
		\State $StageList(i)[Architecture] \gets Best(Stage2, Energy);$
		\State $StageList(i-1)[Architecture] \gets Best(Stage1, Energy);$
		\State $Stage1 = Stage2;$
		\State $Stage2 = Array[];$
		\EndFor
	\end{algorithmic}
\end{algorithm}
\noindent Therefore, our methodology does not effectively evaluate all points to find a pareto-optimal design.

The next phase starts by considering the next stage from \textit{StageList}, and iterates over the inverted lists of the approximation parameters, ranging from least-to-highest approximation.
Since the maximum error magnitude has already been achieved in the previous stage, iterating over the default lists increases the number of explored points and might effectively lead to the same design.
Similar to the first phase, the approximation parameters are used to construct a behavioral model of the current stage in order to evaluate the output quality.
However, since we are traversing from the lower end of the approximation spectrum, we only continue if we find any potential for further maximizing energy reductions.
That means, we store the design only if the quality constraint is satisfied, or we break the loop and move to the next phase of the methodology (lines~\ref{Marker2}$-$\ref{Marker2b}).

Although we have identified the approximation parameters for the two stages currently under consideration \{$i, (i-1)$\}, there is a possibility that we have missed out on another design offering better energy reductions compared to the current designs as we have not explored the higher end approximation spectrum of the second stage.
We address this in the third and final phase of the methodology by traversing diagonally across the number of LSBs approximated, i.e., we reduce the number of LSBs approximated by $2$ for the $(i-1)^{th}$  stage and increase the number of LSBs approximated by $2$ for the $(i)^{th}$ stage.
We reconstruct the behavioral models for these designs and evaluate their output quality. 
If the quality constraint is satisfied, the new designs are stored as a potential approximate processing unit in the two arrays.
This is repeated until the number of LSBs approximated for the $(i-1)^{th}$ stage becomes zero.
We then evaluate the energy reductions of each design in the array, to determine the one which offers the maximum energy reductions.
This design is considered to be the best approximate unit for the current stage.
The second and third phases are continuously repeated until all stages in \textit{StageList} are considered for approximation (lines \ref{Marker3}$-$\ref{Marker3b}).

\noindent 
\section{Experimental Setup}
\label{sec:ES}

\begin{table*}[t]
	\centering
	\vspace{0.1cm}
	\captionsetup{justification=centering, singlelinecheck=false}
	\caption{\textbf{PSNR and Energy Reductions of the Designs Obtained Using the Design Space Generation\\ Methodology for the Pan-Tompkins Data Pre-processing Stage.}}
	\includegraphics[width = \linewidth]{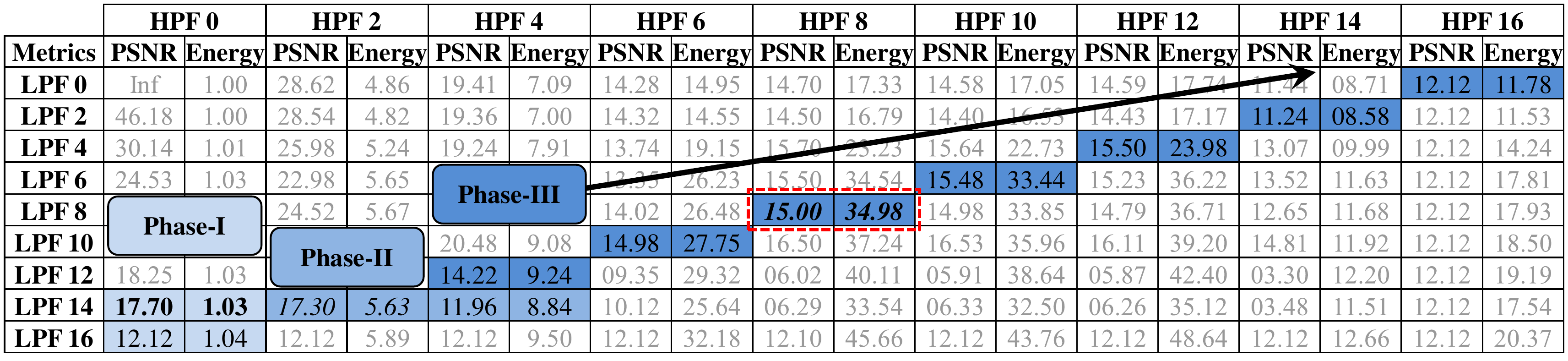}
	\label{tab:LPFHPF}
\end{table*}

\noindent Fig.~\ref{fig:ES} presents an overview of our experimental tool-flow.
The RTL models (implemented in VHDL) of the different approximate adders (32-bit) and multipliers ($16\times16$) along with the five stages (FIR filters) in the Pan-Tompkins algorithm are synthesized using the Synopsys Design Compiler ASIC tool-flow for a 65nm technology library.
We generate detailed area, power, latency, and energy reports for analyzing the resource utilization of the proposed approximate processing units which are then compared with respect to the accurate design.
We also implement behavioral models of the elementary arithmetic modules in MATLAB, so that they can be deployed in the target application (Pan-Tompkins QRS Peak Detection) for quality evaluation purposes.
The output signal observed after the approximated High Pass Filtering, is compared with the accurate output to denote the quality loss, either in terms of PSNR or SSIM, to ensure the output quality of the intermediate signal. 
The final output of the target application needs to be considered while approximating the arithmetic blocks present in the different application stages.
The number of peaks detected in the sample duration, or the peak detection accuracy is the final metric used for quality evaluation in our ECG case study.
We use the MIT-BIH Normal Sinus Rhythm Database (NSRDB) made available online via PhysionetDB~\cite{goldberger2000physiobank}.

\setcounter{figure}{8}
\begin{figure}[h]
	\centering
	\captionsetup{justification=raggedright,singlelinecheck=false}
	\includegraphics[width = \linewidth]{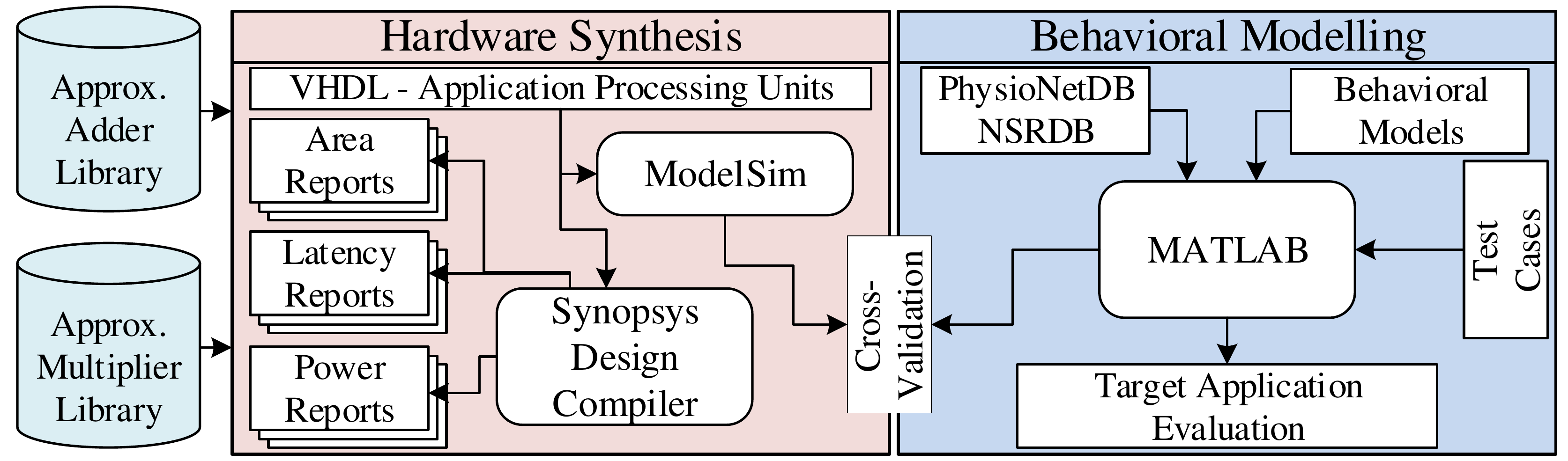}
	\caption{\textbf{Overview of Our Experimental Setup.}}
	\label{fig:ES}
\end{figure}
\section{Results \& Discussion}
\label{sec:Results}

\begin{figure}[h]
	\centering
	\captionsetup{justification=raggedright,singlelinecheck=false}
	\includegraphics[width = \linewidth]{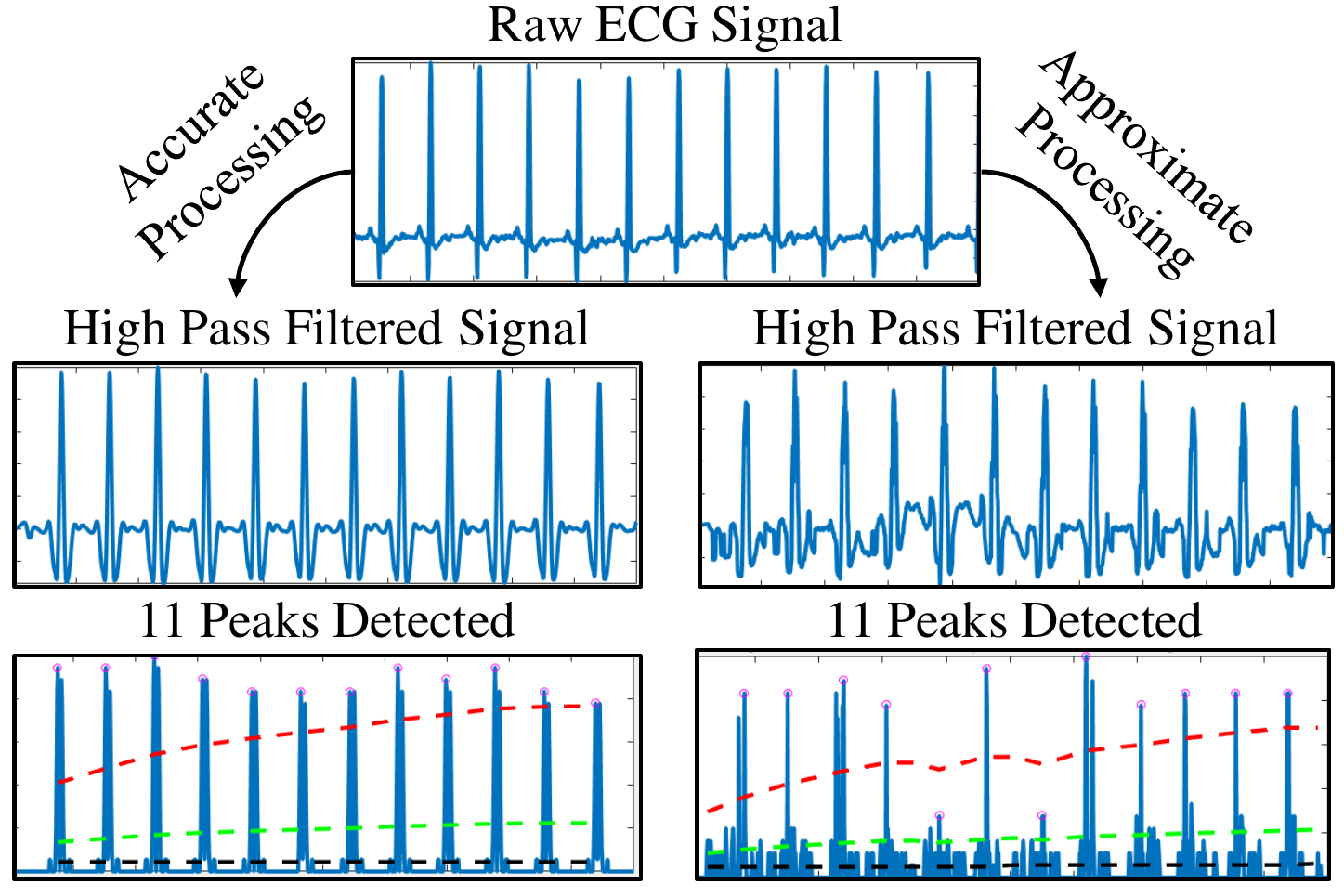}
	\caption{\textbf{Differences in Output Quality Between Accurate and Approximate Processing Units in Pan-Tompkins.}}
	\label{fig:VisualResults}
\end{figure}

\noindent In this section, we illustrate the effectiveness of \textit{XBioSiP}, by applying it on the Pan-Tompkins QRS peak detection algorithm.
Before moving on to illustrating the benefits of our proposed methodology, we first discuss the differences in output signal quality when using accurate and approximate processing units in the Pan-Tompkins Algorithm.
Fig.~\ref{fig:VisualResults} illustrates the differences in signal quality when processing the ECG data-points using accurate and approximate arithmetic blocks with 4 LSBs approximated at all five stages.
Considering the accurate High Pass Filtered signal as a reference, the approximated signal has a PSNR of $19.24$, with $100\%$ peak detection accuracy for the sample duration.
This approximate design requires  \textasciitilde$7\times$ less energy as compared to the accurate counterpart, which significantly increases the lifetime of battery-operated IoT edge devices.

\subsection{Approximations in Data Pre-Processing}
\label{subsec:ASFResults}

\noindent In this subsection, we evaluate the effectiveness of the approximations deployed in the Low Pass and High Pass Filtering Stages of the Pan-Tompkins Peak Detection Algorithm.
For the sake of simplicity, we currently restrict the design space of adders and multipliers to ApproxAdd5 and AppMultV1, respectively~\cite{gupta2013low}\cite{rehman2016architectural}.
We also performed an exhaustive exploration of all $9\times9 = 81$ possible combinations. The corresponding results are presented in Table~\ref{tab:LPFHPF}.
Each simulation consists of an ECG recording that is obtained from the MIT-BIH Normal Sinus Rhythm Database (NSRDB)~\cite{goldberger2000physiobank}.
An ECG recording of $20,000$ samples takes around 300 seconds for filtering and processing. Therefore, an exhaustive exploration of $81$ possible scenarios takes roughly seven hours. 
On the other hand, our design generation methodology successfully generates and evaluates only $11$ designs (in \textasciitilde$1$ hour), where five satisfy the quality constraint.
We considered a PSNR value of $15$ as the user-defined quality constraint for the signal filtering stage.
Out of all possible options, we choose the design which offers the maximum energy reductions of \textasciitilde$35\times$, which obviously satisfies the PSNR value as shown in Table~\ref{tab:LPFHPF}.

\setcounter{figure}{10}
\begin{figure}[b]
	\centering
	\captionsetup{justification=raggedright,singlelinecheck=false}
	\includegraphics[width = \linewidth]{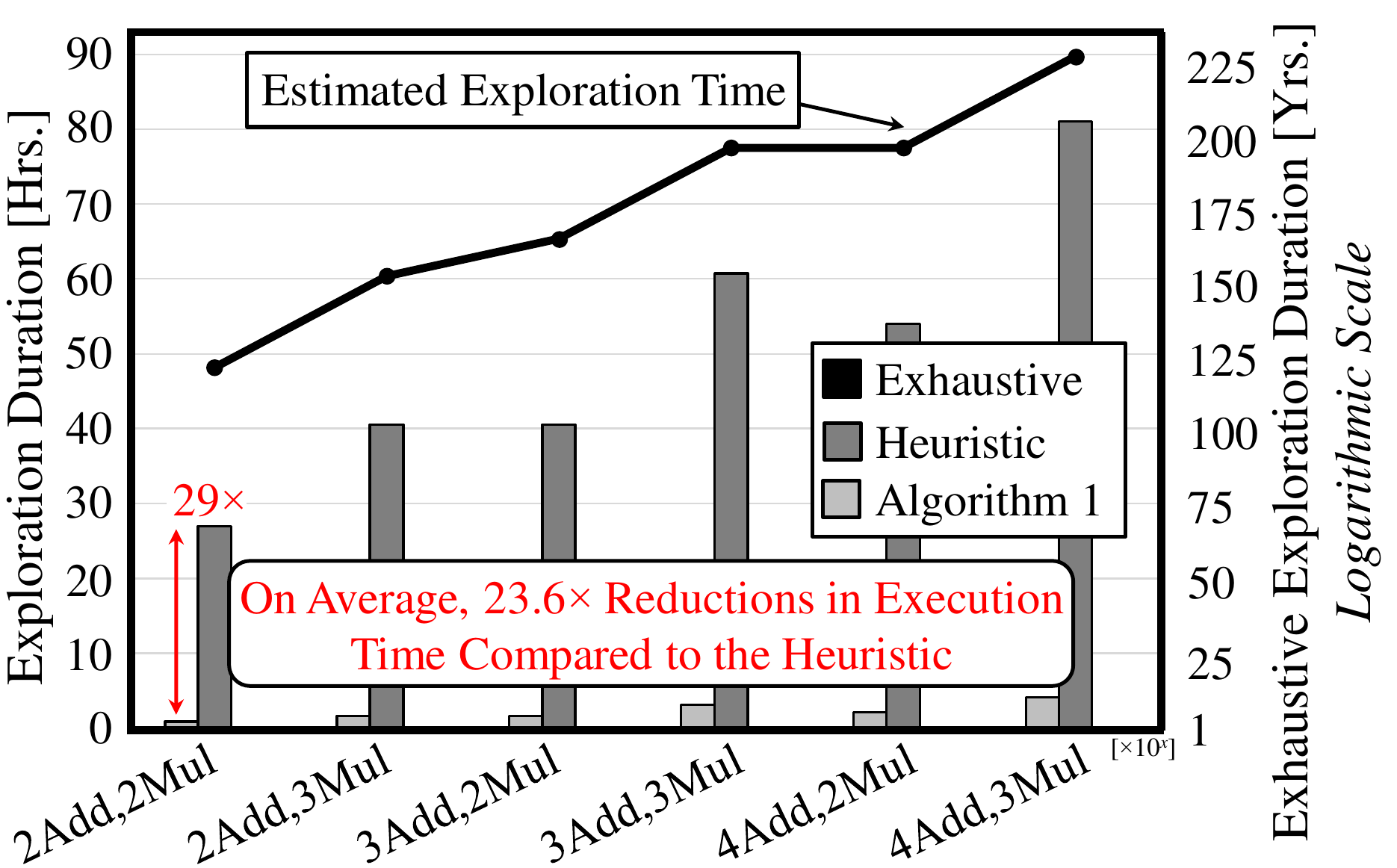}
	\caption{\textbf{Exploration Time Analysis of Algorithm 1.}}
	\label{fig:HeuristicResults}
\end{figure}

\changed{Next, we evaluate the effectiveness of our approach (algorithm~\ref{Algo1}) in reducing the execution time of the design space search by deploying it on the data pre-processing stage of our target application (LPF and HPF).
The exhaustive search involves varying all three major parameters, namely,
\begin{inlinelist}
    \item number of LSBs approximated (0 to 16),
    \item number of approximate adders (AccAdd to AppAdd5), and
    \item number of approximate multipliers (AccMult to AppMultV2),
\end{inlinelist}
to explore the entire design space and identify the design offering maximum energy savings while satisfying the signal quality constraint.
To reduce the overall size of the design space, we implement thresholds and limitations on all the major parameters such as utilizing the same elementary approximate adder and multiplier module throughout the entire design, which is dubbed as the heuristic.
This also includes reducing the LSBs approximated to multiples of $2$.
Fig.~\ref{fig:HeuristicResults} illustrates the execution time analysis of our proposed approach compared with the baselines such as exhaustive and heuristic.
As can be observed, we reduced the exploration time by \textasciitilde$23.6\times$, on average, when compared to the heuristic baseline.}

\subsection{Approximations in Signal Processing}
\label{subsec:ASPResults}

\setcounter{figure}{11}
\begin{figure}[b]
	\centering
	\captionsetup{justification=raggedright,singlelinecheck=false}
	\includegraphics[width = \linewidth]{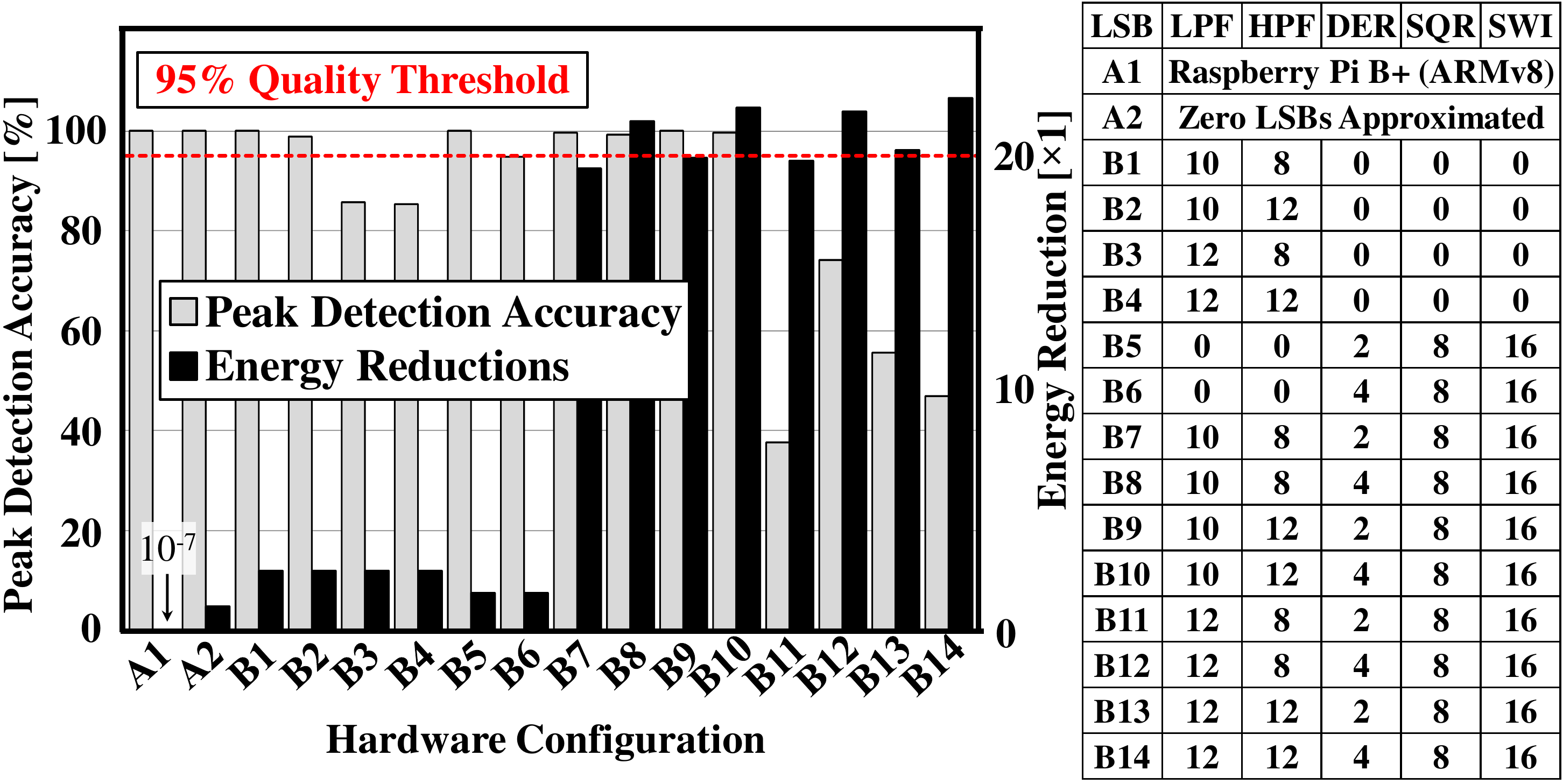}
	\caption{\textbf{Energy-Quality Evaluation of the Approximate Designs Proposed for the Pan-Tompkins Algorithm.}}
	\label{fig:ResultsEnergy}
\end{figure}

\setcounter{figure}{12}
\begin{figure*}[t]
	\centering
	\captionsetup{justification=raggedright,singlelinecheck=false}
	\includegraphics[width = \linewidth]{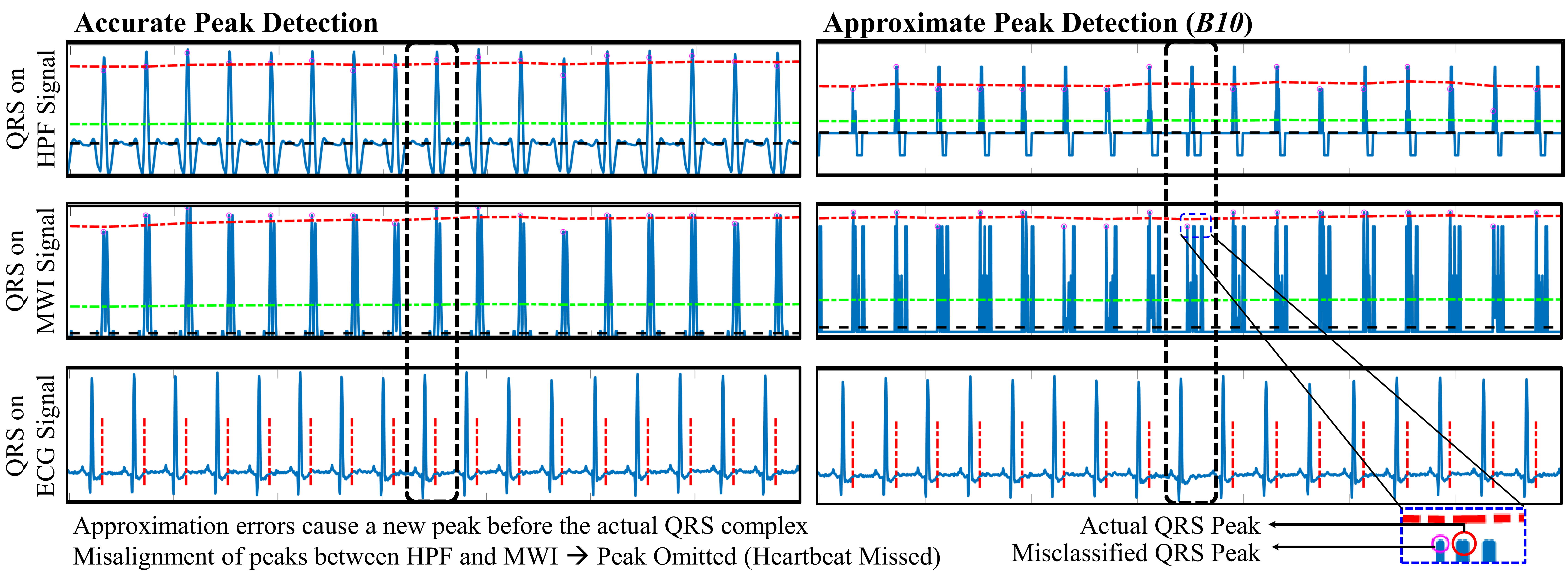}
	\caption{\textbf{Heartbeat Misclassification Analysis of the Approximate Processing Unit (B10).}}
	\label{fig:ErrorAnalysis}
\end{figure*}

\noindent Considering $0\%$ quality loss during the data pre-processing stage, we evaluate the output quality by deploying approximations in the signal processing stages.
We restrict the design space by limiting the number of approximable LSBs to 4, 8, and 16, for the differentiator, squarer, and moving average stages, respectively.
Similar to the previous stage, for the sake of simplicity, we restrict the list of elementary approximate adders and multipliers to ApproxAdd5 and AppMultV1.
We exhaustively evaluated all $135$ possible combinations to extract energy reductions and accuracy values for each design (in~\textasciitilde$11$ hours).
Our proposed design generation methodology is equally applicable in this case to reduce the exploration time and generate design points that offer high energy reductions.
We obtain two Pareto-optimal points from the design space by extracting the Pareto-frontier. Similarly, we extract four Pareto-optimal designs for the data pre-processing stage. The output quality and energy results of these Pareto-optimal designs are presented in Fig.~\ref{fig:ResultsEnergy}.
We also perform a design space exploration of the designs obtained from the data pre-processing and signal filtering stages to analyze the percentage loss in output quality and obtained energy reductions of an end-to-end system.
The results of this design space exploration are presented in Fig.~\ref{fig:ResultsEnergy}.
\changed{The energy reductions presented are obtained with respect to the accurate hardware design with zero approximations (denoted by \textit{A2}).
We also evaluate the energy consumption of the application by executing it on the Raspberry Pi 3 B+ (ARM v8) with HDMI and WiFi switched off (denoted by \textit{A1}).
The energy consumption of \textit{A1} is \textasciitilde$7$ orders of magnitude higher than the energy consumption of \textit{A2}.}
Design \textit{B9} reduces the energy consumption by \textasciitilde$19.7\times$ while detecting all the peaks present in the database.
Design \textit{B10} reduces the energy consumption by \textasciitilde$22\times$ while tolerating a loss of $1\%$ in peak detection accuracy.

We also analyze the output signal of the design \textit{B10} to understand why less than $1\%$ heartbeats were missed by comparing it with the output of \textit{A2}.
Fig.~\ref{fig:ErrorAnalysis} illustrates the differences in output signal of the accurate (\textit{A2}) and approximate processing units (\textit{B10}).
The errors introduced by the approximate arithmetic blocks cause the algorithm to misclassify the error as a peak.
Due to the misalignment of peaks between the HPF and MWI signals (larger than a preset threshold), the detected peak is omitted as an error in classification, and the heartbeat is missed.

\section{Conclusion}
\noindent We presented a novel approximate bio-signal processing methodology, \textit{XBioSiP}, for achieving energy reductions in energy-constrained, sensory, IoT edge devices, and wearable electronics.
After compiling a library of the elementary approximate modules, we evaluate the error-resilience of all application stages to determine the upper-bound of the approximation parameters.
Then we use our proposed design generation methodology to develop approximate stages of the target application that satisfy the quality constraint.
We propose to evaluate the quality constraint twice, after an intermediate stage (data pre-processing) as well the final stage (signal processing), to ensure fine-grained quality control of the intermediate signal.
We evaluate the effectiveness of \textit{XBioSiP} using an ECG processing application called the Pan-Tompkins Algorithm.
We have successfully reduced the energy consumption by \textasciitilde$19.7\times$ for $0\%$ loss in peak detection accuracy, and by \textasciitilde$22\times$ for less than $1\%$ quality loss.
Furthermore, to enable further research and development in this field, we have open-sourced the behavioral and RTL implementations of our approximate modules at \textcolor{blue}{\url{https://xbiosip.sourceforge.io/}}.
In the future, we plan to extend our work to include diagnostic techniques and algorithms across multiple bio-signal processing domains such as ECG-based arrhythmia detection and EEG-based seizure prediction.

\section*{Acknowledgements}
The authors would like to thank Florian Kriebel for his feedback and detailed technical comments, which helped us improve the quality of this work.

\bibliographystyle{ACM-Reference-Format}
\bibliography{sample-bibliography}

\end{document}